\documentclass[aps,prd,twocolumn,superscriptaddress,showpacs,amsmath,amssymb,amsthm,nofootinbib, preprintnumbers]{revtex4-2}
 \usepackage{amsmath}
\usepackage{graphicx}
\usepackage{epstopdf}
\usepackage{float}
\usepackage{hyperref}
\usepackage{color}
\usepackage[T1]{fontenc}
\usepackage[utf8]{inputenc}
\usepackage[toc,page]{appendix}
\usepackage[usenames,dvipsnames]{xcolor}
\usepackage[normalem]{ulem}
\usepackage{lipsum, babel}
\usepackage[justification=justified,format=plain]{caption}

\usepackage{enumitem}   
\usepackage{subcaption}
\usepackage[utf8]{inputenc}
\usepackage{url}
\usepackage{xcolor}
\usepackage{physics}

\usepackage{hyperref}

\usepackage{graphicx}% Include figure files
\usepackage{dcolumn}% Align table columns on decimal point
\usepackage{bm}% bold math
\newcommand{\be}{\begin{equation}}             %:skip:
\newcommand{\ee}{\end{equation}}               %:skip:
\newcommand{\ba}{\begin{eqnarray}}
\newcommand{\ea}{\end{eqnarray}}

\begin{document}

\title{Thermodynamics of four-dimensional regular black holes with an infinite tower of regularized curvature corrections}

\author{Adolfo Cisterna}
\email{adolfo.cisterna@mff.cuni.cz}
\affiliation{Sede Esmeralda, Universidad de Tarapac{\'a}, Avenida Luis Emilio Recabarren 2477, Iquique, Chile}
\affiliation{Institute of Theoretical Physics, Faculty of Mathematics and Physics,
Charles University, V Holešovickách 2, 180 00 Prague 8, Czech Republic}

\author{Mokhtar Hassaine}
\email{mokhtar.hassaine@gmail.com}
\affiliation{Instituto de Matem\'atica, Universidad de Talca, Casilla 747, Talca, Chile}

\author{Ulises Herandez-Vera}
\email{uhernandez.vera@gmail.com}
\affiliation{Instituto de Matem\'atica, Universidad de Talca, Casilla 747, Talca, Chile}

\date{\today}          
\begin{abstract}

We study the thermodynamics of a class of four-dimensional black hole solutions arising from the compactification of a higher-curvature gravity theory featuring an infinite tower of Lovelock-type invariants. For planar horizons, we identify two distinct branches: a regular black hole supported by a nontrivial scalar field and a non-regular general relativity (GR) solution with a trivial scalar profile. Despite their differing geometries, both branches share the same free energy at fixed temperature, revealing a thermodynamic degeneracy naturally linked to the enhanced symmetry and scale invariance of the planar base manifold. In the case of a spherical horizon, even if the scalarized branch is not obtained in closed form, one can see that the degeneracy persists in the absence of the quadratic curvature contribution. On the other hand, if this quadratic term is taken into account, the regular solution may be  thermodynamically favored (or not) over the Schwarzschild–AdS solution depending on the values of the coupling constants of the theory. 
\\

\end{abstract}

\maketitle

%%%%%%%%%%%%%%%%%%%%%%%%%%%%
\section{Introduction}
%%%%%%%%%%%%%%%%%%%%%%%%%%%%

Singularity theorems in General Relativity (GR) demonstrate that, under fairly general physical conditions—such as the validity of energy conditions and the existence of trapped surfaces—gravitational collapse leads inevitably to the formation of spacetime singularities \cite{Penrose:1964wq,Hawking:1967ju,Hawking:1970zqf}. These singularities are characterized by the divergence of curvature invariants and geodesic incompleteness. They represent regions where the predictive power of GR breaks down, and thus signal a fundamental limitation of the classical theory in the strong gravity regime.\\

%The appearance of curvature singularities is widely interpreted as an indication that GR is incomplete at high energy scales. Near these singularities, quantum gravitational effects are expected to become significant, necessitating a more fundamental framework. Although a complete theory of quantum gravity remains elusive, the construction of modified gravitational models that admit non-singular black hole solutions—geometries that remain regular everywhere, including at the center of the collapse, the so-called regular black holes—has received significant attention. 

The appearance of curvature singularities is widely interpreted as an indication that GR is incomplete at high energy scales. In the vicinity of such singularities, quantum gravitational effects are expected to become significant, pointing to the need for a more fundamental theoretical framework. Although a complete theory of quantum gravity remains elusive, considerable attention has been devoted to constructing modified gravitational models that admit non-singular black hole solutions—geometries that remain regular everywhere, including at the center of collapse. These are commonly referred to as regular black holes. For a comprehensive overview, we refer the reader to the review \cite{Lan:2023cvz} and the references cited therein.\\

These solutions, by construction, avoid inner curvature singularities while preserving essential features of black hole spacetimes, such as event horizons. Unfortunately, in most of the cases, they require the inclusion of exotic matter fields and undesired tunings between the theory and solution parameters. \\

It has long been understood that in spherically symmetric cases, any regular black hole geometry must feature a de Sitter-like core near the origin $r=0$, where the curvature invariants remain finite and geodesic completeness is ensured. This necessary and sufficient condition is known as the Sakharov criterion \cite{Sakharov:1966aja,Chinaglia:2018uol}. The aforementioned inner de Sitter region effectively replaces the singular core of standard black hole solutions, such as those found in the Schwarzschild or Reissner-Nordström metrics. The first regular solution derived from equations stemming from an action principle was constructed some time ago in the context of nonlinear electrodynamics \cite{Ayon-Beato:1998hmi}. Within scalar-tensor theories, regular solutions have also been found for certain classes of DHOST theories by means of the Kerr-Schild construction \cite{Babichev:2020qpr, Baake:2021jzv}. In addition, in the case of hairy solutions belonging to the beyond-Horndeski class, regular spacetimes have been obtained under a specific relation between the mass and the hair parameter \cite{Bakopoulos:2023fmv, Baake:2023zsq}.\\

More recently, a promising approach to constructing such regular black holes has emerged through the use of higher-curvature theories of gravity, particularly the class known as quasitopological gravities \cite{Oliva:2010eb,Myers:2010ru,Dehghani:2011vu,Ahmed:2017jod,Cisterna:2017umf}. These theories generalize Einstein gravity by including specific combinations of curvature invariants that contribute non-trivially to the field equations in dimensions higher than four, while still yielding second-order equations of motion under the assumptions of spherical symmetry. In addition, in contrast to Lovelock theory \cite{Lovelock:1971yv,Lovelock:1970zsf}, they are shown to exist at any order in the curvature in dimensions greater than four \cite{Bueno:2019ycr,Bueno:2022res,Moreno:2023rfl}, and even more to provide a basis of polynomial densities for any aimed effective-field-theory expansion of GR \cite{Bueno:2019ltp}.

By employing the full tower of quasitopological gravity terms at arbitrary order $n\geq3$ in the curvature tensor—along with the 
$n=2$ Gauss-Bonnet term—it becomes possible to construct black hole solutions that are regular at all points in the spacetime (of at least dimension five), in vacuum and with only mild assumptions on the theory couplings \cite{Bueno:2024dgm}. Taking the limit $n\rightarrow\infty$ naturally leads to the emergence of a de Sitter core, while the infinite series of higher-curvature terms also enables the construction of various regular geometries, depending on the chosen parametrization of the couplings that govern each term in the series.
These solutions generically contain an inner (Cauchy) horizon, which is typically subjected to mass inflation and other instabilities \cite{Poisson:1989zz}. However, such instabilities may be avoided if the inner horizon structure is extremal \cite{DiFilippo:2024mwm}.\\

Due to the intrinsically higher dimensional nature of quasitopological gravities, the resulting regular black holes inherently reside in higher dimensions. The presence of the full quasitopological tower relies on the structure of gravity theories that are well-defined only in dimensions greater than four. Consequently, any solution constructed in this setting must be interpreted as higher-dimensional, which poses a conceptual challenge: how can such regularization mechanisms, which depend fundamentally on higher-dimensional structures, be applied in a four-dimensional context?\\

This question has been addressed in recent work by supplementing Einstein's theory with four-dimensional Lovelock-like corrections at all orders in the curvature \cite{Fernandes:2025fnz}. These corrections are achieved via a well-defined dimensional regularization of the entire Lovelock tower of higher curvature invariants, that considers the subtraction of two conformally related Lovelock actions of order $n$ and performs the limit to the critical dimension $d=2n$. This is made in strict analogy with the recently explored conformal regularizations of Einstein-Gauss-Bonnet gravity in dimension four \cite{Gurses:2020ofy,Fernandes:2020nbq,Hennigar:2020lsl,Lu:2020iav}. 
Effectively,  the higher-dimensional information is encoded through a scalar degree of freedom. This scalar field arises naturally from the conformal regularization scheme and couples to gravity in a way that retains second-order field equations, thereby avoiding issues related to higher-derivative instabilities or ghosts \cite{Ostrogradksi,Woodard:2015zca}. In fact, it reproduces Horndeski gravity \cite{Horndeski:1974wa}, in a Covariant Galileons fashion \cite{Deffayet:2009wt}, with fixed theory potentials\footnote{This theory was obtained for the first time in  \cite{Colleaux:2020wfv}.}. \\

Importantly, for a planar base manifold geometry\footnote{Although the spherical base manifold counterpart of this solution remains elusive in exact form, below we provide evidence supporting the loss of regularity in this case.}, the resulting black hole solutions in this reduced theory remain everywhere regular, mirroring the higher-dimensional configurations of quasitopological gravity. Again, considering the entire set of corrections allows one to effectively create a de Sitter core. Furthermore, the scalar field introduced in the process does not require any form of self-tuning or fine-tuning to maintain regularity. Instead, the regular geometry emerges as a consequence of the structure of the theory itself. \\

This work is devoted to the study of the thermodynamic properties of the regular planar Anti-de Sitter (AdS) black holes constructed in \cite{Fernandes:2025fnz}\footnote{Thermodynamic analysis of the aforementioned vacuum higher dimensional solutions \cite{Bueno:2024dgm} have been recently addressed in \cite{Aguayo:2025xfi,Hennigar:2025ftm}.}. Using the Euclidean action approach \cite{Gibbons:1976ue}, we compute the mass, entropy, and temperature, and verify that these quantities satisfy the first law of thermodynamics. Interestingly, we find that these thermodynamic charges are degenerate: the vacuum planar AdS black hole in pure GR possesses the same mass, entropy, and temperature, rendering it thermodynamically indistinguishable from the regular black hole. This degeneracy is attributed to the planar geometry of the horizon and has been observed previously \cite{Hennigar:2020fkv} in a three-dimensional non-rotating setup\footnote{In fact, universality of the thermodynamic quantities is no longer valid if rotation is included \cite{Hennigar:2020drx}.}. 
Taking advantage of the planar symmetry, we follow the methodology of \cite{Horowitz:1998ha} to construct the corresponding planar AdS soliton. We compute its mass and free energy, enabling an analysis of the phase transition structure between the soliton and the regular black hole configurations. 
Finally, although the exact form of the spherical counterpart of the regular (hairy) black hole is not known, we compute its mass, entropy, and temperature. In contrast to the planar case, we find that the thermodynamic degeneracy with the Schwarzschild-AdS black hole is lifted. We conclude with a qualitative analysis of the corresponding phase transition structure in this setup.

%This work is devoted to the study of the thermodynamic properties of the regular planar Anti-de Sitter (AdS) black holes constructed in \cite{Fernandes:2025fnz}. By means of the Euclidean action approach \cite{} we compute mass, entropy, and temperature, verifying the first law of thermodynamics. In addition, we observe that these quantities are degenerated, in the sense that the vacuum planar AdS black hole of pure GR does possess the same thermodynamic charges, and therefore is thermodynamically neutral with respect to the regular black hole. This behavior is attributed to the planar geometry of the horizon, and it has been observed before \cite{}. Taking benefit of the planar structure, we construct, along the lines of \cite{}, the corresponding planar AdS soliton and compute its mass and free energy, allowing the study of the phase transition structure between both hairy configurations. Subsequently, we also explore these thermodynamic aspects in the light of the extended thermodynamic scheme in which the cosmological constant is upgraded to be a thermodynamic variable. Finally, even though the spherical counterpart of the hairy black hole is not known in exact form, we compute its corresponding mass, entropy and temperature, showing that in this case there is no thermodynamic degeneracy with respect to the vacuum Schwarzschild AdS black hole. We provide a qualitatively analysis of the phase transition structure of this case. 

%%%%%%%%%%%%%%%%%%%%%%%%%%%%%%%%%%%%%%%%%%%%%%%%%%%%%%%%
\section{Black hole thermodynamics}

\subsection{Thermodynamics of the regular planar black holes}
%%%%%%%%%%%%%%%%%%%%%%%%%%%%%%%%%%%%%%%%%%%%%%%%%%%%%%%

As discussed in \cite{Fernandes:2025fnz}, the conformal dimensional regularization of Lovelock theory gives rise to a scalar-tensor theory with second-order field equations. It is therefore natural to recast this target theory within the Horndeski gravity framework \cite{Horndeski:1974wa,Deffayet:2009wt}. In fact, to order $n$ in the Lovelock Lagrangian $\mathcal{L}^n$, the resulting scalar-tensor theory is described by the Horndeski Lagrangian $\mathcal{L}_H^{(n)}$:
\begin{equation}
\begin{aligned}
&\mathcal{L}_H^{(n)} = G_2^{(n)} - G_3^{(n)}\Box\phi + G_4^{(n)}R
\\
&+ G_{4X}^{(n)} \left[(\Box\phi)^2 - \left(\nabla\mu \nabla_\nu \phi\right)^2\right]
+ G_5^{(n)} G^{\mu\nu}\nabla_\mu \nabla_\nu \phi\\
&- \frac{G_{5X}^{(n)}}{6} \left[
(\Box\phi)^3 - 3\Box\phi \left(\nabla_\mu \nabla_\nu \phi\right)^2
+ 2\left(\nabla_\mu \nabla_\nu \phi\right)^3
\right],
\end{aligned}
\label{HorndeskLagr}
\end{equation}
with fixed theory potentials given by
\begin{equation}
\begin{aligned}
&G_2^{(n)} = 2^{n+1}(n-1)(2n-3) X^n,\\
&G_3^{(n)} = -2^{n} n(2n-3) X^{n-1},\\
&G_4^{(n)} = 2^{n-1} n X^{n-1},\\
&G_5^{(n)} = -\begin{cases} 4 \log X, \qquad n=2, \\ 2^{n-1} \dfrac{n (n-1)}{n-2} X^{n-2}, \qquad n > 2, \end{cases}
\label{Horndeskicouplingfunctions}
\end{aligned}
\end{equation}
where $X = -\dfrac{1}{2} \nabla_\mu \phi \nabla^\mu \phi$ denotes the standard canonical kinetic term of the scalar degree of freedom \cite{Colleaux:2020wfv}.\\

Therefore, our guiding action principle is defined by the four-dimensional action
\begin{equation}
S = \int \mathrm{d}^4 x \sqrt{-g} \left[ R - 2\Lambda + \dfrac{1}{l^2} \sum_{n=2}^{\infty} l^{2n} \mathcal{L}_H^{(n)} \right],
\label{eq:action}
\end{equation}
where $l$ denotes a new length scale that governs the scalar-tensor trace left by the original Lovelock invariants.
%, and where the cosmological constant $\Lambda$ will later be defined in terms of the (A)dS radius $L^{-2}=-\Lambda/3$. 
We have also already defined a set of couplings, that later on we denote by $c_n$, that already provide the planar regular black hole of \cite{Fernandes:2025fnz}. \\

We proceed with the study of the thermodynamics of the solutions using the Euclidean approach, where the thermodynamic ensemble corresponds to the Euclidean path integral in the saddle-point approximation around the classical Euclidean configuration \cite{Gibbons:1976ue}. Given the static and spherically symmetric nature of the geometries, we consider the Euclidean class of metrics
\begin{align}
ds^2 &= N(r)^2 f(r) d\tau^2 + \dfrac{dr^2}{f(r)} + r^2 (dx_1^2 + dx_2^2),\\
\phi(r) &= \ln(r),
\end{align}
with Euclidean time $0 < \tau < \beta$, where $\beta = T^{-1}$ is, as usual, the inverse of the temperature. For concreteness, we assume the on-shell form of the scalar field $\phi(r) = \ln(r)$, which can indeed be derived from the mini-superspace.
%, as shown in the Appendix.

After straightforward computations, the Euclidean action becomes
\begin{align}
I_E &= -\vert\Omega_2\vert \beta \int N \dfrac{d}{dr} \left[ 2r^3 \left( \sum_{k=1}^{\infty} \dfrac{1}{l^2} \left( \dfrac{-l^2 f}{r^2} \right)^k - \dfrac{\Lambda}{3} \right) \right] dr \nonumber\\
&\quad + B_E,
\end{align}
where $B_E$ is a boundary term to be fixed in order to have a well-defined variational principle, and where $\vert\Omega_2\vert=\int dx_1 dx_2$. It is clear that variation with respect to the lapse function $N$ yields the following \textit{polynomial} equation for $f$
\begin{eqnarray}
\dfrac{2r^3}{l^2} \sum_{k=1}^{\infty} \left( \dfrac{-l^2 f(r)}{r^2} \right)^k = \dfrac{2\Lambda r^3}{3} + M,
\label{relation}
\end{eqnarray}
where $M$ is an integration constant. This solution corresponds to the regular black hole configuration recently constructed in \cite{Fernandes:2025fnz}. 

On the other hand, from the extremal condition $\delta I_E = 0$, the variation of the boundary term proceeds as $\delta B_E = \delta B_E(\infty) - \delta B_E(r_h)$. We first compute the variation at infinity, obtaining
\begin{align}
\delta B_E(\infty) &= \vert\Omega_2\vert\beta \lim_{r\to\infty} \dfrac{\delta f}{l^2} \left[ \sum_{k=2}^{\infty} \dfrac{2k(-l^2)^k f^{k-1}}{r^{2k-3}} - 2rl^2 \right], \nonumber\\
&= \vert\Omega_2\vert\beta \delta M.
\end{align}
Hence, we get
\begin{equation}
B_E(\infty) = \vert\Omega_2\vert\beta M = \beta \vert\Omega_2\vert \left( -\dfrac{2 r_h^3 \Lambda}{3} \right),
\end{equation}
where we used relation \eqref{relation}. As usual, at the horizon $\delta f_{r_h} = -4\pi T \delta r_h$, and thus
\begin{align}
\delta B_E(r_h) &= \vert\Omega_2\vert\beta \dfrac{\delta f_{r_h}}{l^2} \left[ \sum_{k=2}^{\infty} \dfrac{2k(-l^2)^k f^{k-1}}{r^{2k-3}} - 2r l^2 \right]_{r_h}, \nonumber\\
&= 8\pi \vert\Omega_2\vert r_h \delta r_h.
\end{align}
Integrating yields
\begin{equation}
B_E(r_h) = 4\pi \vert\Omega_2\vert r_h^2,
\end{equation}
allowing us to express the final form of the boundary term as
\begin{equation}
B_E = \beta \vert\Omega_2\vert \left( -\dfrac{2 r_h^3 \Lambda}{3} \right) - 4\pi r_h^2\vert\Omega_2\vert.
\end{equation}
It is then straightforward to extract the mass $\mathcal{M}$ and entropy $\mathcal{S}$
\begin{eqnarray}
\mathcal{M} = -\dfrac{2 r_h^3 \Lambda}{3}\vert\Omega_2\vert, \qquad \mathcal{S} = 4\pi  r_h^2\vert\Omega_2\vert.
\end{eqnarray}
The temperature can be obtained by differentiating relation \eqref{relation} and evaluating at $r = r_h$
\begin{equation}
T = -\dfrac{r_h \Lambda}{4\pi}.
\end{equation}
Finally, one can directly verify the first law of black hole thermodynamics
\begin{eqnarray}
d\mathcal{M} = T d\mathcal{S}.
\end{eqnarray}
Note that, for the same underlying theory, there also exists a solution corresponding to that of GR with a constant (trivial) scalar field. However, this GR branch is singular in the planar case, singular in the sense it contains a central singularity. Despite differences in the geometries and scalar field profiles between the non-regular GR solution and the regular scalarized configuration, it is straightforward to show that both possess the same free energy at fixed temperature. This thermodynamic equivalence indicates a degeneracy between the two branches, suggesting that the scalar field in the planar case can be activated without affecting the on-shell action—effectively regularizing the geometry without altering its thermodynamic properties.\\

Now, it is possible to re-introduce the coupling constants $c_n$ in front of each of the Horndeski Lagrangians ${\cal L}^{(n)}_H$  \eqref{eq:action}. The action principle we consider is then given by \cite{Fernandes:2025fnz}
\begin{equation}
S = \int \mathrm{d}^4 x \sqrt{-g} \left[ R - 2\Lambda + \dfrac{1}{l^2} \sum_{n=2}^{\infty} c_n l^{2n} \mathcal{L}_H^{(n)} \right],
\label{eq:actioncn}
\end{equation}
and, in this case, the Wheeler relation \eqref{relation} becomes slightly modified as 
\begin{equation}
\dfrac{2r^3}{l^2} \sum_{k=2}^{\infty} c_k\left( \dfrac{-l^2 f(r)}{r^2} \right)^k-2f(r) r = \dfrac{2\Lambda r^3}{3} + M.
\end{equation}
Since the contribution of the variation at infinity depends only on the constant of integration $M$, and at the horizon only on the Einstein-Hilbert term $-2f(r)r$, it is evident that the resulting thermodynamic quantities are insensitive to the specific values of the coupling constants $c_n$. Consequently, for different choices of these couplings (provided they at least ensure the convergence of the series), the thermodynamics remains unchanged. This highlights once again the remarkable nature of these regular solutions: their thermodynamic behavior is universal, unaffected by the detailed structure of the higher-order terms in the action. This robustness suggests that the physical properties of these configurations are governed primarily by the leading-order dynamics, reinforcing their special role within the broader solution space of the theory.

%%%%%%%%%%%%%%%%%%%%%%%%%%%%%%%%%%%%%%%%%%%%
\subsection{Regular  Soliton: The phase diagram of regular black holes}
%%%%%%%%%%%%%%%%%%%%%%%%%%%%%%%%%%%%%%%%%%%%

In pure GR, the Schwarzschild-AdS black hole with a spherical horizon exhibits a rich phase transition structure relative to the maximally symmetric AdS background. As shown by Hawking and Page \cite{Hawking:1982dh}, there exists a critical temperature $T_c$ at which the large-radius Schwarzschild-AdS black hole becomes thermodynamically favored over thermal AdS. This marks a first-order phase transition which, from the perspective of the dual conformal field theory (CFT), corresponds to a confinement/deconfinement transition \cite{Witten:1998zw}.

It is well recognized that the phase transition structure—on both the gravitational and field theory sides—depends on the topology of the spacetime. In fact, no Hawking-Page transition occurs for planar Schwarzschild-AdS black holes with respect to thermal AdS, as the planar black hole is always thermodynamically dominant for any non-zero temperature.

However, even a minimal change in the topology on which the field theory resides—such as compactifying one of the planar directions into a circle—introduces a negative Casimir energy associated with the resulting non-trivial topology. The corresponding geometry on the gravity side is known as the Horowitz-Myers AdS soliton \cite{Horowitz:1998ha}. As a matter of fact, a Hawking-Page transition does exist in between the planar Schwarzschild-AdS black hole and the AdS soliton \cite{Surya:2001vj}.

Owing to the planar topology of the black hole geometry, this soliton can be obtained simply by performing a double Wick rotation: one between the Lorentzian time coordinate $t = ix_1$ and, let us say, the coordinate $x_1 = it$, where we assume that $x_1$ is compactified as $0 < x_1 < L_1$. Without any loss of generality, we can assume that the remaining planar direction satisfies $0\leq x_2<L_2$. Hence, the soliton metric in our case reads 
\begin{equation}
ds^2=-r^2 dt^2+\frac{dr^2}{f(r)}+f(r)dx_1^2+r^2dx_2^2, 
\end{equation}
where $f(r)$ corresponds to the regular black hole solution \eqref{relation}. Now, for this class of regular planar black holes, the Euclidean action $I_{E,s}$, defined with a Euclidean time $\tau=it$ such that  $0\leq \tau<\beta_s$, reads
\begin{equation}
  I_{E,s}=-\beta_s \frac{8\vert\Omega_2\vert }{3\Lambda^2}=-\beta_s \frac{8 L_1L_2}{3\Lambda^2}, 
\end{equation}
from where it is direct to obtain the mass of the soliton
\begin{equation}
{\cal M}_{s}=-\frac{8 \vert \Omega_2\vert}{3\Lambda^2}=-\frac{8 L_1L_2}{3\Lambda^2}.   
\end{equation}
On the other hand, the entropy of the soliton can be computed via the Cardy-like formula introduced in \cite{BravoGaete:2017dso}, using the very same soliton mass. It yields  
\begin{equation}
{\cal S}=3\pi \left(-2{\cal M}_{s}\right)^{\frac{1}{3}} {\cal M}^{\frac{2}{3}}, 
\end{equation}
result in agreement with the one obtained via the standard thermodynamical relation $S_s=I_{E,s}-\beta_s\frac{\partial I_{E,s}}{\partial\beta_s}$.\\ 

Now, having the soliton geometry as the real vacuum state, which has a negative energy, allows the study of phase transition with respect to the planar black hole geometry. In fact, the difference of free energy between both hairy configurations is 
\begin{eqnarray}
    \Delta F=\beta^{-1}\left[I_E-I_{E,s}\right]=\vert\Omega_2\vert \frac{8}{3\Lambda^2}\left[1-\frac{8\pi^3}{\beta^3}\right].
\end{eqnarray}
Recall that the soliton has no associated temperature, hence in order to compare the corresponding free energy we consider the equilibrium condition $\beta_s=\beta$.
It is direct to observe the emergence of a critical temperature 
\begin{equation}
    T_c=\frac{1}{2\pi},
\end{equation}
at which a first-order phase transition occurs. In addition, once again, it can be proven that the thermodynamic structure is degenerated with respect to the one of the vacuum GR case. 

%\subsection{Extended thermodynamics}
%.....
%%%%%%%%%%%%%%%%%%%%%%%%%%%%%%%%%%%%%%%%%%%%%%%%%%%%%%
\subsection{Spherical case and non regular solution}
%%%%%%%%%%%%%%%%%%%%%%%%%%%%%%%%%%%%%%%%%%%%%%%%%%%%%%

An analogous calculation to the one performed above reveals that the function 
$f$ of a line element with a spherical base manifold does not satisfy a Wheeler-like polynomial relation, such as the one given in equation \eqref{relation}. Although a closed-form expression for the metric function remains elusive, it is still possible to compute the associated thermodynamic quantities. These quantities can be derived and analyzed despite the lack of an explicit metric solution, they read
\begin{eqnarray}
&&{\cal M}=-\frac{8\pi r_h^3\Lambda}{3}+8\pi r_h,\quad T=\frac{r_h(1-r_h^2\Lambda)}{4\pi(r_h^2+2)},\nonumber\\
&&{\cal S}=16 \pi^2  r_h^2+64\pi^2\ln(r_h).
\label{nonregtherm}
\end{eqnarray}
Again, it is straightforward to verify the validity of the first law in this case. By introducing coupling functions in front of the Horndeski Lagrangians \eqref{eq:actioncn}, one finds that only the coupling $c_2$ appears in the thermodynamic relations. Indeed, we obtain
\begin{eqnarray}
&&{\cal M}=-\frac{8\pi r_h^3\Lambda}{3}+8\pi r_h,\quad T=\frac{r_h(1-r_h^2\Lambda)}{4\pi(r_h^2+2c_2)},\nonumber\\
&&{\cal S}=16 \pi^2  r_h^2+64c_2\pi^2\ln(r_h).
\label{thermononregc2}
\end{eqnarray}
This result is not surprising, as we had already observed that the presence of the Gauss-Bonnet coupling leads to a logarithmic contribution to the entropy, see e. g. \cite{Babichev:2022awg}.

In the spherical case, the thermodynamic picture changes significantly compared to the planar scenario. Although the explicit form of the scalarized (hairy) black hole solution is not known, one can still compute its mass, entropy, and temperature, and thereby evaluate its free energy. When comparing with the Schwarzschild-AdS solution of general relativity—which corresponds to a trivial (constant) scalar field—an interesting dependence on the coupling $c_2$ emerges.

Specifically, when $c_2\not=0$, the difference in free energy  can be either strictly positive or strictly negative, depending on the value of $c_2$ and the cosmological constant $\Lambda$. In such cases, either the scalarized (hairy) black hole or the GR branch can be thermodynamically favored at fixed temperature. This behavior points to the presence of a richer phase structure, controlled by the interplay between scalar couplings and background curvature.

On the other hand, when $c_2=0$, the situation changes. Although the free energy difference vanishes and the thermodynamic becomes degenerated with respect to the pure GR spherical case, the scalarized solution is still no longer regular near the origin. In particular, it lacks a de Sitter core and does not behave as $f\sim 1$ near $r=0$, indicating a pathological or singular geometry in that limit.

%%%%%%%%%%%%%%%%%%%%%%%%%%%%%%%
\section{Conclusions}
%%%%%%%%%%%%%%%%%%%%%%%%%%%%%%%

Using the Euclidean formalism, we analyzed the thermodynamics of a regular black hole solution in four-dimensional gravity, derived from an infinite tower of Horndeski theory with a planar base manifold. Our results reveal a marked contrast between black hole solutions with planar and spherical horizons within the same scalar-tensor framework. In the planar case, we identified two branches: a regular black hole supported by a nontrivial scalar field and a singular, GR-like solution with a trivial scalar configuration. Both share the same free energy, indicating thermodynamic degeneracy. This stems from the enhanced symmetry and scale invariance of planar horizons, which permit nontrivial scalar fields to regularize the geometry without affecting the on-shell action. Such behavior is reminiscent of flat-horizon holographic models, where scalar hair can be thermodynamically neutral.\\

%..... comments on extended thermodynamics...\\

In contrast, this degeneracy is lifted in the spherical case. Although the explicit metric for the scalarized black hole is unavailable, we computed its mass, entropy, and temperature, allowing evaluation of its free energy. The resulting difference from the GR branch suggests that the scalar field induces a new thermodynamic phase—similar to spontaneous scalarization in scalar-Gauss-Bonnet and other extended gravity theories, where curvature activates scalar fields and shifts the preferred thermodynamic solution \cite{Doneva:2022ewd}. These findings highlight the influence of horizon topology and base curvature on solution structure, thermodynamics, and possible phase transitions.

Further research is needed to assess the regularity and stability of scalarized black holes in the spherical case. A detailed analysis, potentially involving numerical methods, may clarify whether the solution is truly singular or regularized at higher orders. Investigating the dynamical and thermodynamic stability of these solutions, along with their phase structure, could offer deeper insight into the role of scalar fields in black hole physics and their observational implications.

Future work should also test the conjecture that regular black holes with nontrivial scalar fields can exhibit the same free energy as singular GR solutions with constant scalar configurations. Verifying this will likely require numerical exploration of scalarized solutions and their thermodynamic behavior, especially when analytic metrics are inaccessible. It is crucial to determine whether thermodynamic observables alone—such as free energy or entropy—can signal the regularity of a solution. If so, this could serve as a powerful diagnostic tool in modified gravity, enabling the identification of physically viable, regular solutions even in the absence of explicit spacetime metrics.

%%%%%%%%%%%%%%%%%%%%%%%%%%
\section*{Acknowledgments}
%%%%%%%%%%%%%%%%%%%%%%%%%%
A.C. is partially supported by FONDECYT grant 1250318 and by the GA{\v C}R 22-14791S grant from the Czech Science Foundation. MH gratefully acknowledges the University of Paris-Saclay for its warm hospitality during the development of this project. UHV  is partially supported by ANID grant No.  21231297 and would like to thank the University of Talca for funding a research internship at Charles University in Prague, to which he also expresses his sincere gratitude.

\bibliography{apssamp}

\end{document}